\documentclass[a4paper]{jpconf}
\usepackage{graphicx}

\usepackage{units}
\usepackage{color}
\usepackage{url}

\usepackage{lineno}

\begin{document}
\newcommand{\nuc}[2]{$^{#2}\rm #1$}

\newcommand{\fprompt}{F$_{\rm prompt}$}

\newcommand{\bb}[1]{$\rm #1\nu \beta \beta$}
\newcommand{\bbm}[1]{$\rm #1\nu \beta^- \beta^-$}
\newcommand{\bbp}[1]{$\rm #1\nu \beta^+ \beta^+$}
\newcommand{\bbe}[1]{$\rm #1\nu \rm ECEC$}
\newcommand{\bbep}[1]{$\rm #1\nu \rm EC \beta^+$}

\newcommand{\rootcern}{\textsc{Root}}
\newcommand{\gerda}{\textsc{Gerda}}
\newcommand{\largeGERDA}{{LArGe}}
\newcommand{\PI}{\mbox{Phase\,I}}
\newcommand{\PIa}{\mbox{Phase\,Ia}}
\newcommand{\PIb}{\mbox{Phase\,Ib}}
\newcommand{\PIc}{\mbox{Phase\,Ic}}
\newcommand{\PII}{\mbox{Phase\,II}}

\newcommand{\geant}{\textsc{Geant4}}
\newcommand{\mage}{\myacs{MaGe}}
\newcommand{\decayzero}{\textsc{Decay0}}

\newcommand{\nPlus}{\mbox{n$^+$ electrode}}
\newcommand{\pPlus}{\mbox{p$^+$ electrode}}

\newcommand{\AOE}{$A/E$}

\newcommand{\order}[1]{\mbox{$\mathcal{O}$(#1)}}

\newcommand{\mul}[1]{\texttt{multiplicity==#1}}

\newcommand{\pic}[5]{
       \begin{figure}[ht]
       \begin{center}
       \includegraphics[width=#2\textwidth, keepaspectratio, #3]{#1}
       \caption{#5}
       \label{#4}
       \end{center}
       \end{figure}
}

\newcommand{\apic}[5]{
       \begin{figure}[H]
       \begin{center}
       \includegraphics[width=#2\textwidth, keepaspectratio, #3]{#1}
       \caption{#5}
       \label{#4}
       \end{center}
       \end{figure}
}

\newcommand{\sapic}[5]{
       \begin{figure}[P]
       \begin{center}
       \includegraphics[width=#2\textwidth, keepaspectratio, #3]{#1}
       \caption{#5}
       \label{#4}
       \end{center}
       \end{figure}
}

\newcommand{\picwrap}[9]{
       \begin{wrapfigure}{#5}{#6}
       \vspace{#7}
       \begin{center}
       \includegraphics[width=#2\textwidth, keepaspectratio, #3]{#1}
       \caption{#9}
       \label{#4}
       \end{center}
       \vspace{#8}
       \end{wrapfigure}
}

\newcommand{\baseT}[2]{\mbox{$#1\cdot10^{#2}$}}
\newcommand{\baseTsolo}[1]{$10^{#1}$}
\newcommand{\THL}{$T_{\nicefrac{1}{2}}$}

\newcommand{\UBI}{$\rm cts/(kg \cdot yr \cdot keV)$}

\newcommand{\Uflux}{$\rm m^{-2} s^{-1}$}
\newcommand{\Ucpd}{$\rm cts/(kg \cdot d)$}
\newcommand{\Uexpo}{$\rm kg \cdot d$}
\newcommand{\UexpoYear}{$\rm kg \cdot yr$}

\newcommand{\UMWE}{m.w.e.}

\newcommand{\Qbb}{$Q_{\beta\beta}$}

\newcommand{\validate}{\textcolor{blue}{\textit{(validate!!!)}}}

\newcommand{\improve}{\textcolor{blue}{\textit{(improve!!!)}}}

\newcommand{\missing}{\textcolor{red}{\textbf{...!!!...} }}

\newcommand{\quanta}{\textcolor{red}{\textit{(quantitativ?) }}}

\newcommand{\misscite}{\textcolor{red}{[citation!!!]}}

\newcommand{\missref}{\textcolor{red}{[reference!!!]}\ }

\newcommand{\PC}{$N_{\rm peak}$}
\newcommand{\BIC}{$N_{\rm BI}$}
\newcommand{\PAPR}{$R_{\rm p/>p}$}

\newcommand{\PCR}{$R_{\rm peak}$}


\newcommand{\gline}{$\gamma$-line}
\newcommand{\glines}{$\gamma$-lines}

\newcommand{\gray}{$\gamma$-ray}
\newcommand{\grays}{$\gamma$-rays}

\newcommand{\bray}{$\beta$-ray}
\newcommand{\brays}{$\beta$-rays}

\newcommand{\aray}{$\alpha$-ray}
\newcommand{\arays}{$\alpha$-rays}

\newcommand{\betas}{$\beta$'s}


\newcommand{\tab}{{Tab.~}}
\newcommand{\eq}{{Eq.~}}
\newcommand{\fig}{{Fig.~}}
\renewcommand{\sec}{{Sec.~}}
\newcommand{\chap}{{Chap.~}}

 \newcommand{\fn}{\iffalse \fi} 
 \newcommand{\tx}{\iffalse \fi} 
 \newcommand{\txe}{\iffalse \fi} 
 \newcommand{\sr}{\iffalse \fi} 

\title{Backgrounds in the DEAP-3600 Dark Matter Experiment}

\author{B Lehnert for the DEAP-3600 Collaboration}

\address{
Carleton University, Department of Physics, 1125 Colonel By Drive, Ottawa, (ON) K1S 5B6, Canada}

\ead{bjoernlehnert@gmail.com}

\begin{abstract}
The DEAP-3600 experiment, located at \mbox{SNOLAB}, is searching for dark matter with a single phase liquid argon (LAr) target. For a background-free exposure of 3000 kg$\cdot$yr, the projected sensitivity to the spin-independent WIMP-nucleon cross section at 100 GeV/c$^2$ WIMP mass is 10$^{-46}$ cm$^{2}$.

The experimental signature of dark matter interactions is keV-scale argon recoils producing 128 nm LAr scintillation photons which are wavelength shifted and observed by 255 PMTs. 
To reach the large background-free exposure, a combination of careful material selection, passive shielding, active vetoes, fiducialization and pulse shape discrimination (PSD) is used. 
The main concept of the background rejection in DEAP-3600 is the powerful PSD, employing the large difference between fast and slow components of LAr scintillation light. 
The designed background level of DEAP-3600 is less than 0.6 events in a 3000 kg$\cdot$yr exposure.
The experiment was filled in November 2016 and is currently taking dark matter search data. 

\end{abstract}

\section{Introduction}

DEAP-3600 is a single phase liquid argon (LAr) detector located at the SNOLAB underground laboratory in Sudbury, Ontario, Canada. DEAP-3600 is the largest LAr dark matter (DM) experiment built so far and was designed to collect a total fiducial ``background free" exposure of 3000 kg$\cdot$yr within 3 yr of operation. This target dataset would allow to achieve a WIMP exclusion sensitivity of 10$^{-46}$ cm$^{2}$ at 100 GeV/c$^2$ WIMP mass for spin-independent WIMP-nucleon interactions. 

In order to achieve the large ``background free" exposure with $<1$ background events in the WIMP region of interest (ROI) around 16-32~keV$_{\rm ee}$, various hardware design concepts and analysis techniques are employed. In this document, the hardware background mitigation methods along with the overall detector design are outlined. This is followed by the analysis techniques necessary for background reduction. Finally, the first in-situ measurements of various background components within the initial dataset of 4.7~d are presented.

\section{Detector Design}

The LAr target is contained within a spherical 85~cm radius acrylic vessel (AV) capable of holding up to 3600~kg of LAr. For the initial 4.7~d dataset (4.4~live days) presented at TAUP 2017 \cite{thisTalk,DEAP17} the AV was filled with $3322\pm110$~kg. Currently the detector is operated with $\approx 3250$~kg of LAr. After installation of the AV underground, the inner 0.5~mm of the acrylic surface was removed with a sanding robot ``the resurfacer" \cite{DEAPResurfacer}, effectively removing \nuc{Rn}{222} and its daughters which deposit on the surface during exposure to the underground mine air during construction. Especially the long lived \nuc{Pb}{210} (T$_{1/2} = 22.2$~yr) activity is estimated to be reduced by about a factor of 20.

Particle interactions in the LAr produce 128~nm scintillation light which is shifted to visible light in a 3~$\mu$m thick TPB coating on the 9~m$^2$ inside surface of the AV. The visible light is observed by 255 Hamamatsu R5912 HQE PMTs surrounding the target volume at a 50~cm distance. The PMTs are optically coupled to the AV by 255 transparent acrylic light guides (LG). The LGs serve as a passive shield for neutrons, predominantly produced in the borosilicate glass of the PMTs, as well as to hold a temperature gradient between the 87~K LAr temperature and $\approx273$~K temperature at which the PMTs are operated. The gaps between the LGs are filled with high density polyethylene filler material for the same purpose. The inner detector is enclosed inside a 3.4~m diameter stainless steel sphere (SSS), constantly purged with radon-free nitrogen gas. The SSS is submerged in a cylindrical 8~x~8~m  (diameter x height) ultra-pure water tank serving as further passive shielding from outside neutron and gamma radiation. The water tank is instrumented with 48 outward looking PMTs as an active water Cherenkov muon veto. The whole detector setup is located in the ``Cube Hall" of SNOLAB with an overburden of 2100~m (6000~m.w.e.), reducing the atmospheric muon flux by eight orders of magnitude to 0.27~m$^{-2}$d$^{-1}$.

During normal operation, the LAr is hermetically sealed inside the AV and continuously cooled via nitrogen in a cooling coil. During the initial filling, the LAr was passed trough a process system containing a SAES getter and a radon trap to reduce impurities, radon and its daughters inside the target volume. 

Initially, the LAr was filled through a 30~cm diameter 3~m long neck on top of the AV which slightly breaks the otherwise spherical symmetry. At the lower end of the neck, an acrylic flow guide system is installed in order to control the direction and intensity of the gaseous argon (GAr) entering the detector as well as to block the direct line of sight through the neck for neutrons and gammas. In order to have an additional handle of specific background event topologies in this rather complex neck region, it was instrumented with a separate light detection system. Light guiding fibers are wrapped around the outside of the cylindrical acrylic neck, collecting light in this region which is then detected by 4 independent PMTs forming the so called ``neck veto".

\section{Background Mitigation in DEAP-3600}

The main concept of the background rejection in DEAP-3600 is the powerful pulse shape discrimination (PSD), employing the unique scintillation properties of LAr. Upon energy deposition in the LAr, excimers are formed which can either be excited in singlet or triplet states. These excimers de-excite via emission of 128~nm photons whereas from the singlet state these are allowed transitions with a fast 6~ns lifetime and from the triplet states these are forbidden transitions with a longer 1300~ns lifetime. Particle interactions with different linear energy transfer populate these states differently and thus result in different pulse shapes: nuclear recoils (NR) populate dominantly singlet states and electronic recoils (ER) populate dominantly triplet states. For the detection of visible light with the PMTs the time constant of TPB has to be considered and a short detection time window of $t_s = 150$~ns is defined after the trigger occurs. The ratio of short time window $t_s$ and the total time window $t_{all} = 10000$~ns is used as a PSD parameter $\rm F_{prompt} =  t_s / t_{all}$, which can discriminate WIMP-like NR from ER by a factor of \baseTsolo{10} in the WIMP ROI.

\begin{figure}
\begin{center}
\includegraphics[width=0.8\textwidth]{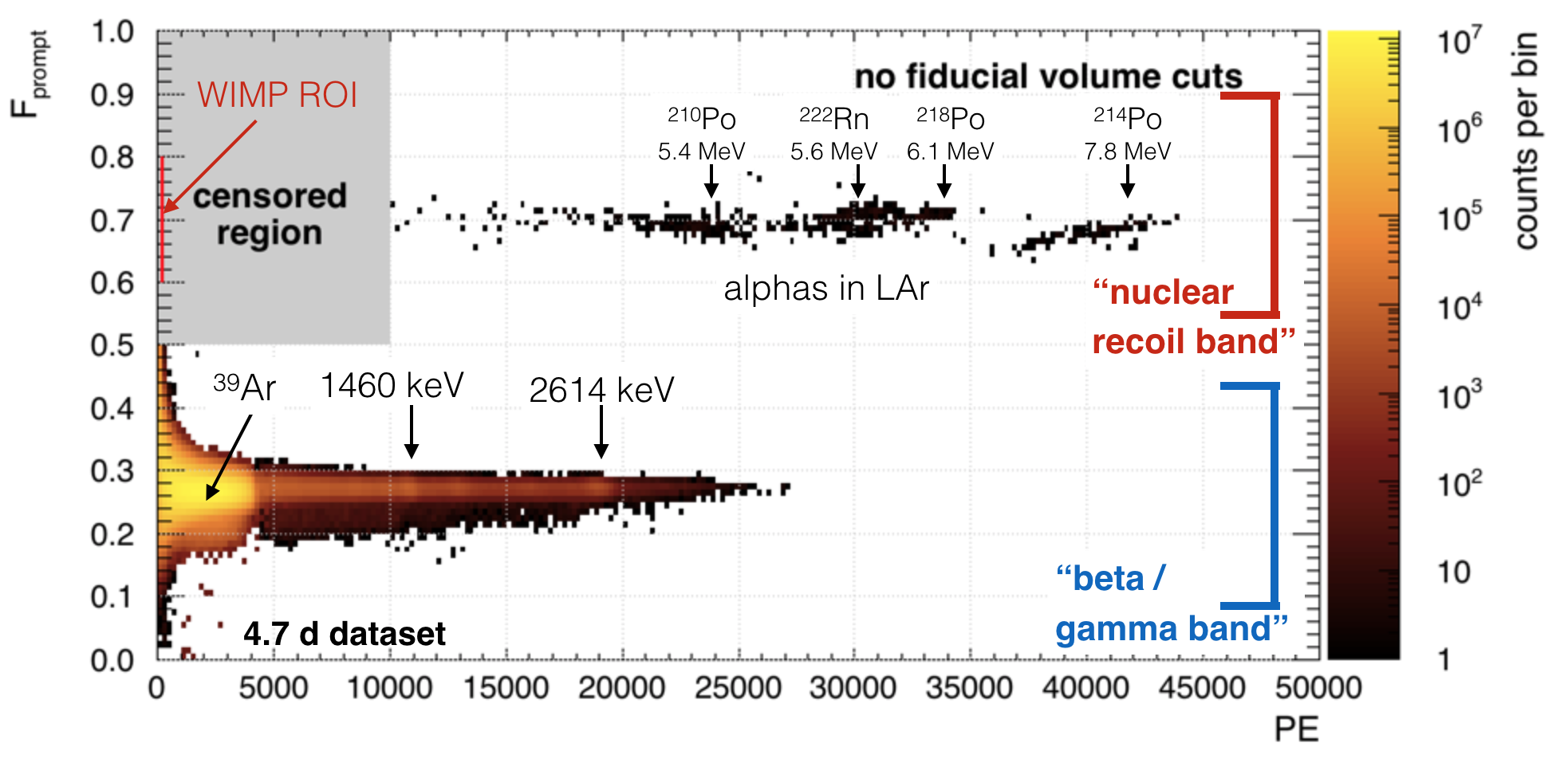}
\end{center}
\caption{\label{fig:fig_fpvspe} PSD parameter (\fprompt) versus detected photo electrons (PE) for the 4.7~d data set. The nuclear recoil and electronic recoil band and their most prominent features are highlighted. The WIMP ROI is at low energies around 120-240 PE. }
\end{figure}

The PSD parameter \fprompt\ is shown versus the detected total photoelectrons (PE) in \fig \ref{fig:fig_fpvspe}. Two bands emerge: At around \fprompt\ = 0.3 for ER interactions and \fprompt\ = 0.7 for NR interactions. Various background populations are highlighted in \fig \ref{fig:fig_fpvspe}: At high \fprompt, alpha decays in the bulk LAr and on the AV surface are observed at high energies from radon and its daughters. The WIMP ROI is located in this band at low energies as indicated. At low \fprompt, ER interactions from LAr internal beta decays and external gamma emitters are observed. The ER band is dominated by \nuc{Ar}{39} decays inside the LAr up to 4500~PE which also dominates the trigger rate in DEAP-3600 with $\approx 3300$~Hz. 

The main background sources in the WIMP ROI are from (1) leakage of high energy alpha decays to low energies due to partial energy depositions in LAr or loss of scintillation light in the neck region, (2) leakage of \nuc{Ar}{39} and other ER interactions from low \fprompt\ to high \fprompt, or (3) neutrons which recoil of \nuc{Ar}{40} nuclei in a similar way that WIMPs do. DEAP-3600 was designed such that the background budget in a 3000 kg$\cdot$yr fiducial exposure dataset would be less than 0.2 events for each of these components, allowing in total less than 0.6 events in the ROI. In the following these background sources are discussed.

\section{Alpha Backgrounds}

The dominant alpha decays in DEAP-3600 are from \nuc{Rn}{222}, \nuc{Rn}{220} and their daughters emanating from bulk materials in the piping and entering the LAr. Another contribution to alpha decays is residual \nuc{Pb}{210} (\nuc{Rn}{222} daughter with T$_{1/2} = 22.2$~yr) on the inner AV surface, decaying into \nuc{Po}{210} which emits alphas of 5304~keV. Alpha decays are a background for WIMP interactions if the majority of light is produced in LAr. GAr alpha interactions show a significantly lower \fprompt\ value and are rejected by PSD. 

Alpha decays with typically $>5$~MeV can reconstruct in the WIMP ROI if (1) the decay occurs on surfaces and only a fraction of their energy is deposited in scintillating material or (2) only a fraction of the scintillation light is detected. In DEAP-3600, case (1) can occur on the LAr-TPB interface, the bulk of the TPB, the TPB-AV interface or in the first microns of the bulk AV material. Alpha decays at these location result in a different energy response of the detector, which is simulated and compared with data. We observe \nuc{Po}{210} decays on the TPB-AV interface with an activity of $0.22\pm0.04$~mBq/m$^2$ on a $7.36$~m$^2$ surfaced covered by LAr. We also see evidence of \nuc{Po}{210} decays in the first 80~$\mu$m thick AV bulk which we constrain to $<2.2$~mBq total activity. Those surface interactions can be effectively removed by a radial fiducial volume cut. Significant shadowing of scintillation light in case (2) can only occur in the neck region inside small pockets of LAr or LAr films. These shadowing effects produce peculiar light distributions in the PMT map and can be discriminated with dedicated topology cuts. In addition, the neck veto is used to remove those events.

Alpha decays in the LAr bulk are observed from \nuc{Rn}{222}, \nuc{Po}{218} and \nuc{Po}{214} as well as from \nuc{Rn}{220}. In the LAr bulk the full event energy is detected; however, at high alpha energies the PMTs and DAQ saturate depending on the radius, and a radial dependent light yield is observed. This explains the slanted alpha peaks in \fig \ref{fig:fig_fpvspe}. Activities are measured with peak counting and time coincidence analyses of \nuc{Rn}{222}$-$\nuc{Po}{218}, \nuc{Rn}{220}$-$\nuc{Po}{216} and \nuc{Bi}{214 }$-$\nuc{Po}{214} tags resulting in \baseT{(1.8\pm0.2)}{-7} Bq/kg, \baseT{(2.0\pm0.2)}{-7} Bq/kg, \baseT{(2.6\pm1.5)}{-9} Bq/kg for  \nuc{Rn}{222}, \nuc{Po}{214} and \nuc{Rn}{220}, respectively. 
There is no evidence for \nuc{Po}{214} collecting on the LAr-TPB interface as was observed in DEAP-1 \cite{DEAP1}. The measured \nuc{Rn}{222} concentration of 0.2~$\mu$Bq/kg in DEAP-3600 is significantly lower than in other current noble liquid dark matter detectors such as PandaX-II (6.6~$\mu$Bq/kg) \cite{alphaPanda}, LUX (66~$\mu$Hz/kg) \cite{alphaLUX} or Xenon-1t (10~$\mu$Bq/kg) \cite{alphaXenon}.

\section{Beta and Gamma Backgrounds}

The ER interactions in a band between between 0.2 and 0.4 in \fprompt\ are dominated by \nuc{Ar}{39} at low energies ($<0.5$~MeV), by \grays\ from external components at medium energies ($0.5-2.6$~MeV) and by \nuc{K}{42} and close \nuc{Tl}{208} sources at high energies ($>2.6$~MeV). 
\nuc{Ar}{39} is a cosmogenic isotope intrinsic in argon extracted from the atmosphere with a specific activity of $1.01\pm0.10$~Bq/kg \cite{Ar39Activity} and a Q-value of $565\pm5$~keV. The dominant sources of external \grays\ are from the PMT glass with specific activities of 139~mBq/kg, 921~mBq/kg and 546~mBq/kg for the \nuc{Th}{232} chain, the \nuc{U}{238} chain and \nuc{K}{40}, respectively. \nuc{K}{42} with a Q-value of $3525.2\pm0.2$~keV is part of a decay chain starting from \nuc{Ar}{42} which is another cosmogenic isotope intrinsic in atmospheric argon with a specific activity around 100~$\mu$Bq/kg \cite{Gerda16,Barabash16}.

\begin{figure}
\begin{center}
\includegraphics[width=0.8\textwidth]{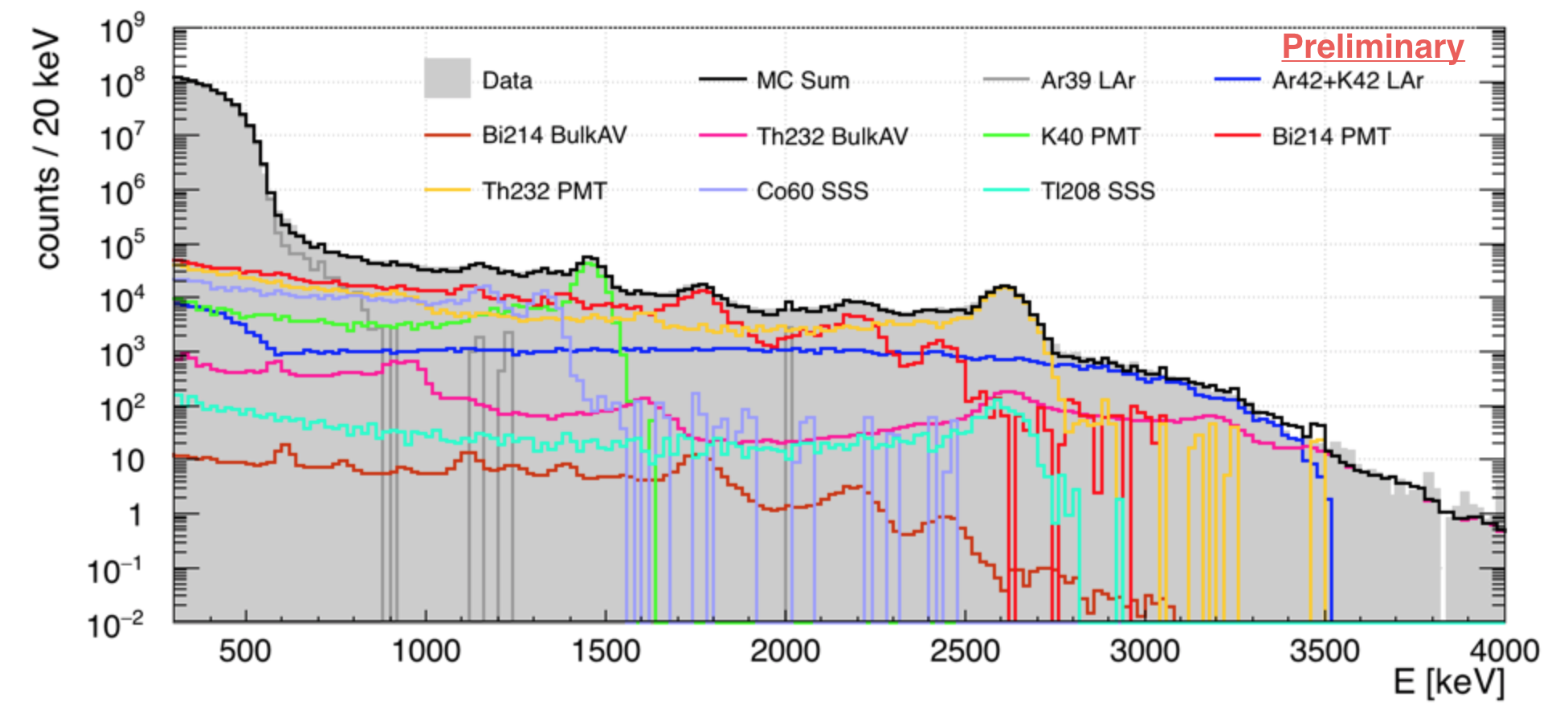}
\end{center}
\caption{\label{fig:fig_ERBand} Electronic recoil background model (back line) in \fprompt\ between 0.2 and 0.4. The individual background contributions (coloured lines) are scale to prior activities from literature values and screening results and to the data set (solid grey). Pile-up and energy resolution are folded into the MC spectra. }
\end{figure}

All relevant beta and gamma emitters were simulated and combined into an ER background model shown in \fig \ref{fig:fig_ERBand}. The grey histogram shows the data in a 9.6~d dataset and the black solid line shows the MC expectation of the model. The coloured solid lines show the individual components of the model. The activity of each component was scaled to the prior knowledge based on screening results and literature values and adjusted within a factor of two to best fit the data. This model should not be considered a fit but rather a proof of principle to illustrate that the ER background in DEAP-3600 can be explained with expected background components over a range of 8 orders of magnitude. The components in the plot comprise \nuc{Ar}{39}, \nuc{Ar}{42} and \nuc{K}{42} inside the bulk LAr, the \nuc{U}{238} and \nuc{Th}{232} chains in the bulk AV, the \nuc{U}{238}, \nuc{Th}{232} chains and \nuc{K}{40} in the PMTs and \nuc{Co}{60} and the \nuc{Th}{232} chain in the SSS. 

The \nuc{Ar}{39} PSD leakage is not discussed in the this document but was found to be a factor of 10 better than expected from a conservative extrapolation based on DEAP-1 \cite{DEAP17}.

\section{Neutron Backgrounds}
The DEAP-3600 design is based on maximally shielding external neutrons with the water tank and acrylic layers, rather than identifying and tagging them. 
Neutrons recoil of \nuc{Ar}{40} nuclei in the same way as WIMPs do and can deposit up to 10\% of their energy in a single elastic scatter. The main sources of neutrons in DEAP-3600 are ($\alpha$,n) reactions induced by natural decay chain alphas in bulk material, fission neutrons and muon induced cosmogenic neutrons. An extensive neutron MC campaign has been performed based on material screening activities and ($\alpha$,n) yields from SOURCES-4C \cite{Wilson02} and NeuCBOT \cite{Westerdale17}. About 70\% of the neutrons and thus the dominant source is expected to originate from ($\alpha$,n) reactions from the \nuc{U}{238} and \nuc{Th}{232} decay chains in the borosilicate PMT glass. The decay chain activities in the PMT glass is strongly constrained by the ER background model through \nuc{Bi}{214} and \nuc{Tl}{208} \grays\  discussed above to an activity within a factor of two of the target value.

A data driven in-situ constraint on the neutron flux is based on the idea that nearly every neutron entering the LAr and producing a NR signal will thermalize and be ultimately captured on \nuc{Ar}{40} or \nuc{H}{1} in the surrounding acrylic. The \nuc{Ar}{40} capture results in a \gray\ cascade with 6.1~MeV whereas the \nuc{H}{1} capture results in a single \gray\ of 2.2~MeV. Searching for coincidences between the initial NR ($>80$ PE) and the subsequent \gray\ interaction results in three observed events within 109~d, which are consistent with the random coincidence background. Considering the detection efficiency of these signals, a limit on neutron interactions is placed which is consistent with expectations; however, the current exposure is insufficient to fully test the expectation with statistical significance. Another in-situ neutron constraint is being developed looking at high energy capture \grays\ which are expected to dominante the ER region $>5.5$~MeV.

\section{Conclusions}

The main expected background sources in DEAP-3600 are alphas, \nuc{Ar}{39} PSD leakage and neutron interactions. 
The bulk LAr alpha activity concentration of \nuc{Rn}{222} daughters has been measured and is more than an order of magnitude lower than other current noble liquid dark matter experiments. The observed \nuc{Po}{210} surface alpha activity has been identified coming from the TPB-AV interface. A contribution coming from the first microns in the AV bulk could be constrained. The collection of LAr-born \nuc{Po}{214} on the LAr-TPB interface was not observed. In summary, the surface alpha contribution to the background in the WIMP ROI is within expectations of the design.
The \nuc{Ar}{39} PSD leakage surpasses expectations by a factor of 10. An electronic recoil background model was developed which can quantitatively describe the ER background up to 3.5~MeV within a factor of two of expectations.  \nuc{U}{238} and \nuc{Th}{232} activities in the PMT glass are the dominant neutron sources through ($\alpha$,n) reactions and are constrained by the ER background model to within a factor of two of expectations. Two methods to directly constrain the neutron flux in-situ are being developed which are, however, not yet sensitive enough to measure the neutron flux.\\
In general, the expected background components are well under control and are refined into quantitative background models. Other potential background sources are currently being investigated such as other light sources e.g.\ from Cherenkov in combination with LAr scintillation events and scintillation events in the neck region.

\section*{Acknowledgments}
This work is supported by the Natural Sciences and Engineering Research Council of Canada, the Canadian Foundation for Innovation (CFI), the Ontario Ministry of Research and Innovation (MRI), and Alberta Advanced Education and Technology (ASRIP), Queen's University, University of Alberta, Carleton University, DGAPA-UNAM (PAPIIT No. IA100316), European Research Council (ERC StG 279980), the UK Science \& Technology Facilities Council (STFC) (ST/K002570/1), the Leverhulme Trust (ECF-20130496). Studentship support by the Rutherford Appleton Laboratory Particle Physics Division, STFC and SEPNet PhD is acknowledged.
We thank SNOLAB and its staff for support through underground space, logistical and technical services. SNOLAB operations are supported by CFI and the Province of Ontario MRI, with underground access provided by Vale at the Creighton mine site.
We thank Compute Canada, Calcul Qu\'ebec and the Center for Advanced Computing at Queen's University for providing the computing resources required to undertake this work.

\end{document}